\documentclass[10pt]{iopart}
\usepackage{iopams,setstack,array}
\usepackage{siunitx,textgreek,bm,braket}
\usepackage{graphicx,dcolumn}
\usepackage[colorlinks=true,urlcolor=magenta,linkcolor=blue,anchorcolor=blue,citecolor=blue]{hyperref}
\usepackage{orcidlink}
\usepackage{newtxtext}
\usepackage[switch]{lineno}
\definecolor{grey}{RGB}{192,192,192}
\newcommand{\minew}[1]{{\color{black}{#1}}}
\newcommand{\miold}[1]{\iffalse{#1}\fi}
\newcommand{\mitip}[1]{\iffalse{#1}\fi}

\begin{document}

\title[PIC simulations of \minew{the }filamentation process in magnetized \minew{RF} plasmas]{Particle-In-Cell simulations of \minew{the }filamentation process in magnetized \minew{radio-frequency }plasma\minew{s} \miold{of capacitively-coupled radio-frequency discharge}}

\author{
    Huidong Huang$^1$\orcidlink{0009-0005-8959-8751},
    Jian Chen$^{1,\footnotemark[1]}$\orcidlink{0000-0001-9807-489X}, and
    Zhibin Wang$^{1,\footnotemark[1]}$\orcidlink{0000-0002-6812-7855}}
\footnotetext[1]{Authors to whom any correspondence should be addressed.}

\address{$^1$ Sino-French Institute of Nuclear Engineering and Technology, Sun Yat-Sen University, Zhuhai 519082, People's Republic of China}
\ead{\url{chenjian5@mail.sysu.edu.cn} and \url{wangzhb8@sysu.edu.cn}}
\vspace{10pt}
\begin{indented}
\item Received July 2024, revised April 2025
\end{indented}

% Abstract & Keywords
\begin{abstract}
In \miold{the}\minew{a} uniform \miold{radio-frequency capacitively-coupled plasma (RF-CCP)}\minew{radio-frequency (RF) plasma} between a large electrode pair, \miold{adding}\minew{the addition of} an axial magnetic field induces diverse longitudinal filaments. \miold{This phenomenon, termed “filamentation”, challenges conventional understanding and remains poorly understood to date.}
To reveal its pattern dynamics, we conduct \minew{two-dimensional (}2D\minew{)} Particle-In-Cell\minew{ (PIC)} simulations\minew{, capturing the entire evolution of the filamentation process.} \miold{to comprehensively examine whole process of filamentation, identifying two distinct stages.}
\minew{%Our results suggest an intrinsic self-organizing mechanism governing the pattern formation across configurations, which exhibits remarkable robustness: variations in lateral boundary types (e.g., open vs. periodic) minimally perturb the fundamental structure of filaments.
We found that the entire evolution experiences two dynamic stages.}
\miold{Initially}\minew{In the first stage}, \minew{electrostatic }standing waves\minew{ and plasma density ripples} grow\miold{s} \minew{synergistically}\miold{with a modulational instability}, forming \miold{regular filaments}\minew{filamentary pattern}.
\minew{Our results show that the plasma ripples and RF electrostatic standing waves are modulated. In addition, each filament equips a double-humped peak. The spectrum reveals that the oscillations are mainly RF and its higher harmonics.}
Subsequently, \miold{when initial wavenumber matching relation breaks, }the plasma shifts towards \minew{a }dynamic regime governed by\minew{ the} competition between Lorentz and thermal pressure forces, characterized by \miold{filaments'}\minew{the} chaotic evolution\minew{ of filaments}.
\minew{Through RF-cycle averaging, our force analysis demonstrated that electrons and ions are governed by the magnetic force and electric force respectively. The time-averaged magnetic force drives electrons to accumulate at plasma density maxima, while time-averaged electric force pushes ions into the same regions, jointly molding the filaments.}
These novel clues pave the way \miold{to}\minew{for a} theoretical\miold{ly} understanding \minew{of}\miold{the} filamentation instability\miold{,} and provide\miold{s} essential references \miold{in}\minew{for} effectively manipulating \miold{the }magnetized plasmas.
%\miold{These novel clues pave the way to theoretically understanding the filamentation instability, and provides essential references in effectively manipulating the magnetized plasmas.}\minew{These novel insights lay the foundation for a theoretical comprehension of filamentation instability and offer crucial references for the effective manipulation of magnetized plasmas.}
\end{abstract}

\vspace{2pc}

\submitto{\PSST}
\maketitle

\ioptwocol

% Main Body
\section{Introduction}\label{Sec1}

Magnetic fields are commonly utilized in \miold{capacitively-coupled plasma (CCP)}\minew{gas} discharges to enhance the plasma density~\cite{zheng_2019_enhancement,zhang_2021_resonant,patil_2022_electron} and modify plasma shape~\cite{barnat_2008_rf,ma_2023_enhancing,dahiya_2023_magnetic}. \minew{Particularly, people have shown a keen interest in magnetizing charged dusty plasmas because of the dust contamination problem in various plasma-related techniques~\cite{fortov_2005_complex,merlino_2021_dusty,melzer_2021_physics}. Two decades ago, Konopka et al. introduced a strong magnetic field into an radio-frequency (RF) discharge chamber, initially aiming to study magnetized dusty plasmas.} \miold{Experiments~\cite{konopka_2005_complex,schwabe_2011_pattern} demonstrate that CCPs}\minew{They discovered that RF plasmas} subject to an axial magnetic field emerge filamentation, giving rise to a sequence of luminous patterns that extends along the magnetic field lines~\cite{konopka_2005_complex}. These patterns can manifest either as organized formations (like concentric rings) or irregularly-moving filaments lacking regularity~\cite{schwabe_2011_pattern,thomas_2019_pattern}\miold{, showcasing diverse morphology that has captured extensive attention~\cite{tsankov_2022_foundations}}. \miold{Remarkably, this phenomenon provides, contrary to common belief~\cite{conrads_2000_plasma}, certain entrance into the bulk plasma for the surface RF disturbance despite lower frequency (e.g., \qty{13.56}{MHz}~\cite{konopka_2005_complex,schwabe_2011_pattern,thomas_2016_initial,thomas_2019_pattern,williams_2022_experimental}) compared to the plasma frequency (several gigahertz).}
\minew{Remarkably, the experiments demonstrated that the filamentation can exhibit without dusts~\cite{konopka_2005_complex,schwabe_2011_pattern,thomas_2019_pattern}, thus it is a phenomenon inherent to pure plasma itself. However, these filamentary patterns can significantly influence the dust motion~\cite{schwabe_2011_pattern,jung_2018_experiments,hall_2020_dynamics}~(see \href{https://bpb-us-e2.wpmucdn.com/wordpress.auburn.edu/dist/3/28/files/2022/10/dust_and_filaments.gif}{video}). This, in turn, offers novel possibilities for the dynamic control of industrial dust.}
\minew{Overall, studying filamentation in magnetized RF plasmas not only enhances our understanding of magnetized plasmas but also has promising implications for a variety of applications, including controlled fusion~\cite{winter_2004_dust,krasheninnikov_2008_recent}, semiconductor etching~\cite{cardinaud_2000_plasma,wu_2010_high}, surface modification~\cite{anders_2005_plasma} and material deposition~\cite{profijt_2011_plasmaassisted}.}
%\minew{It is truly peculiar, and there might be some unfamiliar mechanism underlying it. Therefore, uncovering the physical processes of filamentation is conducive to deepening our understanding of magnetized plasmas.}
\miold{Understanding the fundamental physics of the filamentation is imperative for effectively manipulating the magnetized plasmas, which plays a key role in various CCP techniques such as semiconductor etching~\cite{cardinaud_2000_plasma,wu_2010_high}, surface modified~\cite{anders_2005_plasma} and material deposition~\cite{profijt_2011_plasmaassisted}.}

\miold{For two decades, s}\minew{S}ignificant efforts have been devoted to experimental investigations of \minew{the }filamentary patterns. \miold{When }\minew{Konopka and }Schwabe et al. carried out \miold{preliminary}\minew{the earliest} observations on the formation of luminous filaments~\cite{konopka_2005_complex,schwabe_2011_pattern}\miold{,} \minew{and }they proposed that the filamentation strength depends on the magnetization of ions\minew{. }\miold{, additionally t}\minew{T}heir findings \minew{also }suggested that filamentation modifies dramatically the ion transport~\cite{schwabe_2011_pattern}. Subsequently, Thomas et al. \miold{expansively}\minew{extensively} studied the filamentation in magnetized dusty plasmas~\cite{thomas_2015_observations,thomas_2016_initial,jung_2018_experiments,thomas_2019_pattern}. \miold{Through parameter sweep concerning}\minew{By varying the} magnetic intensity, power deposition, and gas pressure, they \miold{have given}\minew{established} an empirical law \miold{of}\minew{for} the filamentation strength. Recently, highly-resolving \miold{observation}\minew{imaging}~\cite{williams_2022_experimental} has further \miold{identified several modes}\minew{revealed the multi-morphology} of \minew{the pattern }micro-structure\miold{ of filament}, \miold{each occurring under experimental parameters.}\minew{varying with parameter changes.} Results showed that \miold{each filament undergoes}\minew{the filaments undergo} chaotic flowing\miold{ and}\minew{, collective} rotating \miold{ceaselessly}\minew{and spinning (see \href{https://bpb-us-e2.wpmucdn.com/wordpress.auburn.edu/dist/3/28/files/2022/09/MDPX_filaments.gif}{video})}\minew{, showcasing the highly nonlinear filamentation pattern dynamics}. \miold{Overall, experiments over years have revealed increasing complexities, that is, non-linearities and unsteady non-equilibrium state, leaving behind a growing number of problems to be solved.}
%\miold{Overall, \minew{years of }experiments \miold{over years }have revealed \miold{increasing complexities}\minew{the multifaceted complexity of filamentation}, \miold{that is,}\minew{characterized by} \minew{multi-coupled }non-linearities and \miold{unsteady }\minew{chaotic }non-equilibrium \minew{pattern dynamics.}\miold{state, leaving behind a growing number of problems to be solved.}}

%\minew{Complementing experimental efforts, combined numerical–theoretical analysis continues to unravel pattern formation mechanisms in magnetized RF plasmas.}
\minew{Besides experimental efforts, combined numerical–theoretical analysis has also been performed to unravel pattern formation mechanisms in magnetized RF plasmas.}
\miold{Besides the experiments, preliminary numerical studies have also been performed to explore the pattern dynamics in magnetized CCPs.} Menati et al. employed a two-dimensional hybrid model~\cite{kushner_2009_hybrid} to investigate the pattern formation of in a magnetized capacitively-coupled argon plasma \miold{for the configuration with}\minew{between} a \minew{RF-}powered metal electrode and a grounded electrode with a dielectric barrier~\cite{menati_2019_filamentation}. Simulation results showed the dielectricity is strongly correlated with the filamentation strength. Afterwards, \minew{they employed} a three-dimensional (3D) \miold{inertialess} two-fluid model \miold{reproduced}\minew{to reproduce} the longitudinal \miold{concentric stripes}\minew{plasma ripples} akin to experimental observation\minew{s}~\cite{menati_2020_experimentala}. \minew{Based on the simulation results, they considered the filamentary plasma to be generated due to diffusion~\cite{menati_2021_variation}. However, they have recently reinterpreted these structures as Turing patterns~\cite{menati_2023_formation}, which has effectively revealed the magnetic field's the crucial role in driving plasma filamentation.}\miold{They recently progressed to making initial application of linearized fluid model in steady-state assumption and interpreted the pattern phenomenon as merely a higher-order diffusion mode~\cite{chen_2016_introduction}, in omitting both intricacy of real electron dynamics and self-consistency of RF discharge~\cite{menati_2021_variation}.}

\miold{Despite considerable endeavors, decisive theoretical breakthrough for explaining the filamentation is still absent.}\minew{Despite a substantial progress in revealing the mechanism of filamentary pattern formation, the understanding of the dynamics of filamentary plasmas is relatively limited.}
\miold{A central question that remains unsolved is \emph{how magnetic fields facilitate the penetration of RF perturbations into the bulk plasma}. In addition, existing studies~\cite{schwabe_2011_pattern,thomas_2016_initial,thomas_2019_pattern,menati_2019_filamentation,menati_2020_experimentala,menati_2021_variation,williams_2022_experimental} have never well unfolded the detailed process and non-linearities involved in pattern formation. Thus far, another physics question that remains unanswered is \emph{what governs the pattern dynamics}. People have known very little about the dynamical processes involved in the formation and evolution of these patterns, which have impeded in-depth analysis.}
\minew{The primary motivation of this study is to comprehensively characterize the non-equilibrium dynamics of filamentary plasmas. We attempt to provide a kinetic perspective for the entire evolution of filamentation.}\miold{The primary objective of our current work is to address the two major physics questions just posed.}
\miold{Given the inadequacy of available experimental facilities (e.g. International Micro-gravity Plasma Facility (IMPF)~\cite{konopka_2005_complex,schwabe_2011_pattern}, and Magnetized Dusty Plasma Experimental device (MDPX)~\cite{thomas_2012_magnetized,thomas_2016_initial,thomas_2019_pattern}.) in providing the mechanical data required for dynamic analysis, excellent functionality of Particle-In-Cell approaches~\cite{birdsall_1991_particleincell,verboncoeur_2005_particle,arber_2015_contemporary} motivates our interest in launching kinetic simulations for reproducing entire process of filamentation.}

In this \miold{paper}\minew{study}, we conduct simulations of a magnetized argon \miold{RF-CCP}\minew{RF discharge} by employing a Particle-In-Cell/Monte-Carlo-Collision (PIC/MCC) \miold{model}\minew{method}~\cite{vahedi_1993_capacitive,donko_2011_particle,wilczek_2020_electron}. This \miold{model}\minew{simulation technique} incorporates the electron and ion kinetics~\cite{birdsall_1991_particleincell,verboncoeur_2005_particle,arber_2015_contemporary} as well as their \minew{neutral }collisions\miold{ with the background gas}~\cite{vahedi_1995_monte,tskhakaya_2007_particleincell}. \miold{Our simulations comprehensively depict the kinetics involved in the whole process of filamentation.}
%\minew{A comparative analysis of lateral boundary effects suggests that the adopted periodic geometry is optimal for capturing the inherent properties of this process.}
\minew{We segmented the whole process into two dynamic regimes, namely the standing wave growing (SWG) stage and the pattern evolving (PE) stage, and analyzed them separately.}
\miold{We identify two distinct stages therein, that is, firstly the stage of standing wave growing (SWG) and secondly the stage of pattern evolving (PE). During the SWG stage, regular plasma filaments emerge from a uniform background until they reach saturation, providing clews for replying the first question. This stage reveals the presence of modulational instability (MI)~\cite{nishikawa_1968_parametric,nishikawa_1968_parametrica,zakharov_1974_nonlinear,zakharov_2013_nonlinear,gelash_2014_superregular}. In the PE stage, initially regular patterns undergo chaotic evolution, which addresses our second question. The forces governing the pattern dynamics are analyzed through fluid analysis.}
\minew{The multi-timescale analysis characterizes the RF cycle-averaged dynamics of the filamentary plasma.}
\minew{We have identified the key forces that govern the pattern dynamics, providing insight into the maintenance and evolution of filaments.}

This paper is organized as follows: the employed \miold{PIC/MCC model and the numerical settings}\minew{simulation method} are described in section~\ref{Sec2}. The \miold{simulations }results \minew{of the full evolution }are presented and discussed in section~\ref{Sec3}, and a summary is provided in section~\ref{Sec4}. Additional supports are presented in the appendix.
\section{\miold{Simulation Model}\minew{Physical Models and Simulation Method}}\label{Sec2}

\begin{figure*}[htb]
    \centering
    \includegraphics[width=0.75\textwidth]{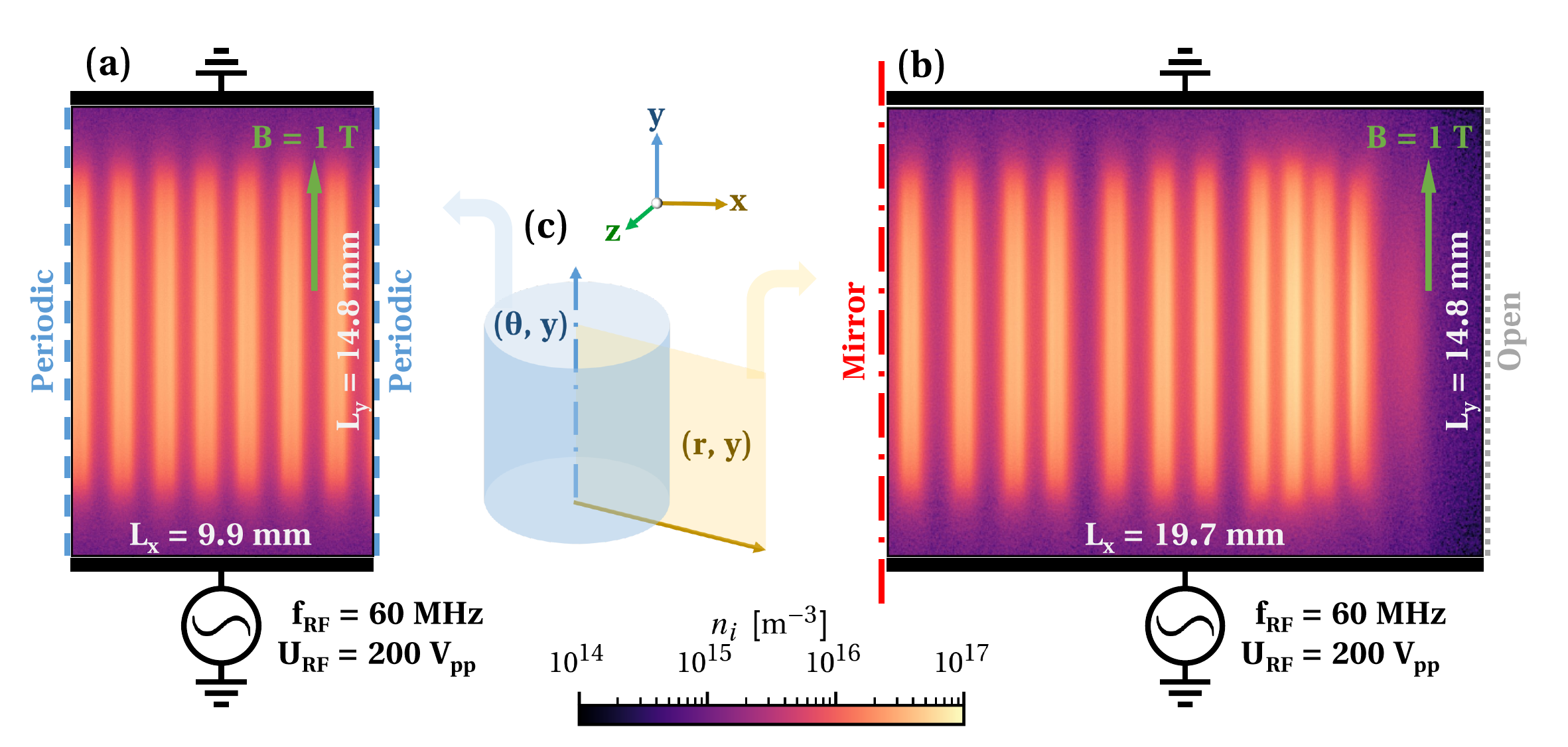}
    \caption{Schematic diagram\minew{s} of the simulation setup\minew{s}.\minew{ The planar simulation domains with dimensions $L_x \times L_y$ and (a) periodic boundaries, as well as (b) open boundaries, are considered to analyze (c) the cross-sections ($\theta$,~$y$) and ($r$,~$y$) respectively. Both cases are at simulation time $t=\qty{0.1}{ms}$, already reached quasi-steady state.}}
    \label{Fig1}
\end{figure*}

\miold{In this study, we adopt a 2D rectangular simulation domain ($L_x \times L_y$) to simulate the filamentation process in magnetized CCPs, as sketched in figure~\ref{Fig1}. It should be noted that some experiments employed a more complex configuration (see IMPF reactor in figure~\ref{Fig6} of \ref{ApxA}). We have performed a test simulation with IMPF-type configuration and the results show a similar pattern formation, justifying the adoption of the configuration in figure~\ref{Fig1}. In this paper, $x$ and $y$ signify separately the radial and axial coordinates respectively.}

% \begin{figure}[htb]
%     \centering
%     \includegraphics[width=1\hsize]{Fig.1.pdf}
%     \caption{Schematic diagram of the setup.}
%     \label{Fig1}
% \end{figure}

\minew{In this study, we adopt two distinct structures of two-dimensional (2D) planar simulation domain ($L_x \times L_y$) with separately (a) periodic boundaries and (b) open boundaries, as sketched in figure~\ref{Fig1}. These two topologies are separately related to two orthogonal cross-sections, ($\theta$,~$y$) and ($r$,~$y$), as shown in figure~\ref{Fig1}(c). Thus $y$ signifies the axial coordinate, and $x$ can represent either the azimuthal ($\theta$) or the radial ($r$) coordinate, with neglecting the limited effects of the geometric curvature, as shown in~\cite{vahedi_1997_simultaneous,ma_2023_enhancing,charoy_2019_2d,villafana_2021_2d}.}

\minew{
    \begin{itemize}
        \item[(a)] Periodic boundaries. It is of size $L_x\times L_y$ = \qtyproduct{9.9 x 14.8}{\mm}. Periodic boundary conditions are used for the both left and right boundaries ($x = 0$ and $x = L_x$). The coordinate domain ($x$,~$y$) should be regarded as the lateral surface ($\theta$,~$y$) of a cylinder.% with a finite radius.

        \item[(b)] Open boundaries. It is of size $L_x\times L_y$ = \qtyproduct[]{19.7 x 14.8}{\mm}. A symmetry axis is defined at the left boundary ($x=0$), while the right boundary ($x=L_x$) is left open. The coordinate domain ($x$,~$y$) could be regarded as an arbitrary longitudinal section ($r$,~$y$) with the axis as one side.
    \end{itemize}

    In both these two topologies, the plasma is immerged in axial external magnetic field ${|\bf B|}={B_y}=\qty{1.0}{T}$; the upper electrode at $y = L_y$ is grounded, and the lower electrode located at $y = 0$ is powered with an RF voltage. The RF voltage provides a sinusoidal waveform $U(t)=U_0 \sin(2\pi f_{\scriptscriptstyle{\text{RF}}} t)$ with $U_0 = \qty{100}{V}$ and $f_{\scriptscriptstyle{\text{RF}}} = \qty{60}{MHz}$. The domain is filled with argon as the background gas and only singly-charged $\text{Ar}^{+}$ are considered in the simulations. Particles that hit these two electrodes are assumed to be absorbed. The vacuum permittivity $\epsilon_0$ is applied to entire space and boundaries. The electrodes with positive permittivity differ from the actual ones, but this numerical setting was justified by \cite{menati_2019_filamentation}.
} %(common metal and ITO materials generally have ${\mathfrak{Re}}(\epsilon_r)<0$ in the RF range~\cite{park_2017_dynamic})

\minew{
    We explored the filamentation process with periodic configuration (figure~\ref{Fig1}(a)) and the open configuration (figure~\ref{Fig1}(b)).
    %While each exhibits distinct characteristics, they share significant structural commonalities.
    %Figure~\ref{Fig1}(b) reveals a plasma sheath forming along the right edge (open boundary), yet notably retains filaments with structural continuity to those in Fig.~\ref{Fig1}(a).
    As the quasi-steady state is reached, both configurations exhibit filaments of comparable structures, magnitudes ($\max\{n_p\}\lesssim\qty{1e17}{m^{-3}}$) and scales (diameter of ${1.5\pm0.1}~\unit{mm}$)\footnote{\minew{In our simulations, the FWHM (full width of half maximum) of plasma density ripple is $\sim\qty{1}{mm}$), which is consistent to the range ($w_\text{fil}\approx\qty{1}{mm}$) reported in the experiments~\cite{konopka_2005_complex,schwabe_2011_pattern,thomas_2019_pattern,williams_2022_experimental}.}}, which suggests boundary conditions (periodic vs. open) impose relatively limited effects on the filament's structure.
    %This comparison reveals that the wavelength selection, in this context, is predominantly driven by certain instability mechanisms, while boundary conditions play a relatively minor role.
    %Alterations in boundary conditions (e.g., periodic vs. open) or spatial scales (tested over \qtyrange{10}{20}{mm} heights) impose minimal effects on fundamental structure of the filaments.
    These observations demonstrate that the filamentation phenomenon exhibits cross-configuration robustness.
    Hence, for simplicity, we select the periodic configuration to illustrate the pattern formation and nonlinear dynamics.

    %This paper investigates the fundamental commonalities inherent to filamentation phenomena. The periodic configuration (figure~\ref{Fig1}(a)) is particularly advantageous for this analysis: Its minimalist geometry and spatial symmetry eliminate extrinsic disturbances in the horizontal dimension. This configuration, in turn, isolates extraneous instabilities and systematically reduces irrelevant variables. These attributes unambiguously position this configuration as an ideal model for revealing the physical essence of filamentation phenomenon.

    %Notably, the $x$-dimension characteristics exhibit universal applicability, governing not merely the azimuthal direction but all orientations orthogonal to the $y$-axis. To emphasize this dimensional generality, we adopt the broader terminology “horizontal”/“transverse” rather than azimuth-specific descriptions. Subsequent analysis will focus primarily on this periodic configuration.
}

\minew{
    In these models, the external circuit is not considered. This is because both directly-coupled~\cite{konopka_2005_complex} and capacitively-coupled~\cite{menati_2019_filamentation,thomas_2019_pattern} discharges have been experimentally demonstrated to exhibit filamentation, i.e., the blocking capacitor is not a necessary complexity for the occurrence of this phenomenon. %Therefore rigorously, though what we simulate is a directly-coupled discharge rather than a capacitively-coupled discharge, this will not disrupt the essence of filamentation. Moreover, this exclude an unnecessary external complexity, which helps us concentrate on the magnetized plasma itself.
}

\miold{
    The system lengths in $x$ and $y$ directions are denoted as $L_x = 9.9~\text{mm}$ and $L_y = 14.8~\text{mm}$, respectively. The lower electrode located at $y = 0$ is powered with a radio-frequency voltage (sinusoidal waveform of amplitude $U_0 = 100~\text{V}$ and working frequency $f_{RF} = 60~\text{MHz}$), and the upper electrode at $y = L_y$ is grounded. Particles that hit these two electrodes are assumed to be absorbed. The secondary electron emission~\cite{horvath_2018_effect} is disabled in the present work. This is justified by the fact that turning on the secondary electron emission has slight influence on the physical process and steady-state plasma parameters (see the results shown in figure~\ref{Fig7} of \ref{ApxA}). Periodic boundary conditions are used for the left and right boundaries ($x = 0$ and $x = L_x$), as shown as vertical dashed lines in figure~\ref{Fig1}. A uniform magnetic field ${|\bf B|}=1.0~\text{T}$ is applied and aligned with the $y$ axis. The domain is filled with argon as the background gas and only singly-charged $\text{Ar}^{+}$ are considered in the simulations. The heat map shows a typical spatial distribution of $\text{Ar}^{+}$ within the 2D simulation domain in current work.
}

All the simulations in this study are performed using a open-source PIC/MCC code named EDIPIC-2D. EDIPIC-2D has been well benchmarked against multiple codes~\cite{charoy_2019_2d} and has been successfully applied to numerous low-temperature plasma simulations, as recently reported in the literature~\cite{janhunen_2018_evolution,sun_2022_physical,xu_2023_rotating}. \minew{Note that although the code is 2D in position space, it is 3D in velocity space.}

\miold{To reduce the computational time of profile buildup process, the}\minew{These} simulation\minew{s} start\miold{s} with a uniform plasma with the plasma density $n_{p,0} = n_{e,0} = n_{i,0} = \qty{2.5E16}{m^{-3}}$, the electron temperature $T_{e,0} = \qty{3.0}{eV}$ and the ion temperature $T_{i,0} = \qty{0.026}{eV}$. The background gas is argon and its pressure is fixed as \qty{6.0}{Pa}. Initially, 500 super-particles per cell for each species are created, with the particle weight $w = 73728$. \minew{After dozens of microseconds, these systems ultimately reaches a quasi-steady state.
%In theory, the stable interval of RF discharge equilibrium is irrelevant to the perturbed parameters. And also, in our simulations, the filamentation gradually emerged after reaching a preliminarily discharge equilibrium.
As exemplified in figure~\ref{Fig6}, regardless of whether the $n_{p,0}$ is set to \qty{1E14}{m^{-3}}, \qty{2.5E16}{m^{-3}} or \qty{1E17}{m^{-3}}, the number of super-particles eventually converges to a narrow range\minew{, suggesting that the minimal effects of initial parameters on the quasi-steady state}.
%In essence, the influence of the initial state on the quasi-steady state is extremely weak.
}

Three types of electron-neutral collisions, \minew{namely,}\miold{i.e. the} elastic scattering, excitation and ionization, as well as \miold{the }$\text{Ar}^{+}$-Ar charge exchange collisions are included.
The cross-section data used in the simulations are taken from~\cite{cramer_1959_elastic,banks_1966_collision,banks_1966_collisiona,sakabe_1991_cross}.
\minew{The secondary electron emission (SEE)~\cite{horvath_2018_effect} is disabled in the present work, because SEE has a limited impact on the filamentation process (see the results shown in figure~\ref{Fig7} of \ref{ApxA}, which are consistent with the findings in~\cite{menati_2019_filamentation}).}

\minew{The spatial grid size and time step used in the simulation were carefully selected. }The cell size $\Delta x =\Delta y = \qty{38.4}{\um}$ is able to capture the Debye wavelength $\lambda_{De}$ and the ion Larmor radius $r_{Li}$ ($r_{Li} = \qty{144}{\um}$ and $\lambda_{De}$ are on the order of $\qty{0.1}{mm}$). \miold{Note that r}\minew{R}esolving electron Larmor radius ($r_{Li} =\miold{\qty{130}{\um}}\minew{\qty{6}{\um}}$) requires a smaller cell size and leads to a unaffordable computation cost.
Since the characteristic width of the filaments (\qty{1.4}{mm}) is much larger than the electron Larmor radius, we neglect the finite \minew{electron }Larmor radius effects~\cite{ramos_2005_fluid,cerri_2013_extended,smolyakov_2017_fluid} on the filamentation process. The time step $\Delta t = \qty{18.7}{ps}$ is certain to resolve the electron plasma frequency \minew{$f_{pe}\lesssim\qty{3}{GHz}$}\miold{$f_{pe} = \qty{2840}{MHz}$} as well as the radio-frequency $f_{RF}=\qty{60}{MHz}$. To simulate the slow motion of filaments, the simulation was run for \qty{1}{ms} (\miold{namely }53\miold{0} million time\minew{-}steps). \miold{To capture the high frequencies physics, 24 probes are equidistantly arranged in 4~columns~$\times$~6~rows at the bottom-left corner to record the time-dependent evolution of density, potential and electric field.}
\section{Simulation Results and Discussions}\label{Sec3}

\minew{
    In our simulations, the RF plasmas exhibit filamentation under a strong axial magnetic field, yet the filaments disappear at \qty{0}{T}. These results align with experimental observations~\cite{schwabe_2011_pattern,jung_2018_experiments} and numerical results~\cite{menati_2019_filamentation}.

    Our simulation results indicate that the entire evolution of plasma filamentation can be subsumed into two developmental phases: the standing wave growing (SWG) stage and the pattern evolving (PE) stage.
    \begin{itemize}
        \item[{\bf I}.] The standing wave growing (SWG) stage encompasses pattern initiation and amplification. In this stage, the transverse RF electrostatic standing wave and plasma filaments emerge from a uniform background.

        \item[{\bf II}.] The pattern evolving (PE) stage represents a chaotic stage during the entire evolution. At this stage, the plasma filaments can flow erratically, merge in pairs or even surge suddenly from depletion region.
    \end{itemize}
    Once the pattern reaches saturation, the filamentary plasma transitions from the SWG stage to the PE stage.

    In the following two sub-sections, we will analyze these two stages separately. In this paper, we aim to provide a panoramic view to the entire filamentation process, rather than elaborating on the RF dynamics details (cf. the supplemental videos~\footnote{\minew{See Supplemental Material [url] for more details on the RF dynamics.}}).
}

\subsection{Standing Wave Growing Stage}\label{Sec3A}

%\minew{During the standing wave growing (SWG) stage, an transverse RF standing wave emerges from a uniform background and intensifies. Meanwhile, the plasma density evolves gradually into multiple rippled structures. These two processes are mutually reinforced and develop synchronously, thus driving the growth of filamentation instability. In our simulations, both the electric field components $E_x$ and $E_y$ display standing waves with multiple-RF frequencies overall. The magnitudes of electric field and plasma density keep increasing until saturation, which marks the end of the SWG stage.}

\begin{figure}[htb]
    \centering
    \includegraphics[width=1\hsize]{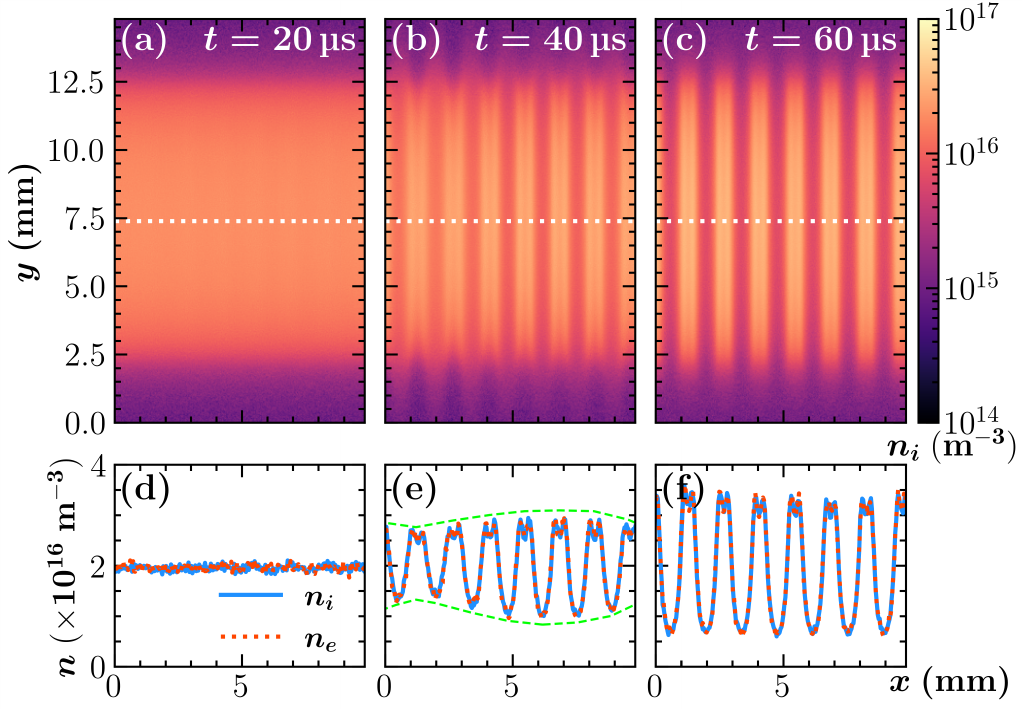}
    \caption{Standing wave growing stage.
    The $\text{Ar}^+$ 2D profiles present the plasma distribution at distinct moments: $t$ = (a)~\qty{20}{\us}, (b)~\qty{40}{\us}, (c)~\qty{60}{\us}. The white dot-dashed lines highlight the half-height altitude $y=\qty{7.4}{mm}$. Blue and red curves indicating respectively $\text{Ar}^+$ and $\text{e}^-$ depicted in (d--f) are their corresponding 1D profiles along the half-height slice. Typical modulational feature manifests on the \miold{spatial standing waveform}\minew{plasma ripples} in (e), which is enveloped by lime dashed curves.
    }
    \label{Fig2}
\end{figure}

Figure~\ref{Fig2} shows the evolution of \minew{the }ion density distributions in \miold{standing wave growing}\minew{the SWG} stage\miold{, which is characterized by a standing wave arising from uniform background and growing until reaching saturation}. \miold{Sub-f}\minew{F}igures~\ref{Fig2}(a--c) depict the ion density distribution at \qtylist{20; 40; 60}{\us}, while figures~\ref{Fig2}(d--f) show the corresponding ion density distribution along the mid-plane ($y=\qty{7.4}{mm}$), denoted by the white lines in figures~\ref{Fig2}(a--c). As shown in figure~\ref{Fig2}(a), ion density keeps relatively uniform in the first \qty{20}{\us} in the bulk region. Two RF sheaths are formed near the upper and lower electrodes with the thickness around 2 mm. The center plasma density is about $\SI{2e16}{\per\cubic\meter}$. Subsequently, \miold{radial}\minew{horizontal} disturbances surge from uniform background, leading to the \miold{density distribution with standing waveform }\minew{rippled structures}, as depicted in figures \ref{Fig2}(b) and \ref{Fig2}(c). These \miold{standing waves }\minew{ripples} maintain coherent structures that resembles the filaments observed in experiments~\cite{schwabe_2011_pattern,thomas_2019_pattern}.

As seen in figures~\ref{Fig2}(d-f), the \miold{amplitude of the standing wave}\minew{magnitude of density perturbation} keeps growing, during which quasi-neutral condition is still fulfilled (electron and ion densities are almost equal). \miold{More i}\minew{I}nterestingly, as depicted in figures \ref{Fig2}(e) and \ref{Fig2}(f), the \miold{wave}\minew{shape} is clearly modulated, \miold{i.e. the amplitudes of crests and troughs at different locations are varied, }exhibiting an envelope structure \miold{as a whole }(denoted by the lime dashed curves). \minew{Note that modulated envelopes exist in the electric field as well. Modulated waveform like this are often associated with parametric instabilities~\cite{nishikawa_1968_parametric,nishikawa_1968_parametrica,zakharov_1972_collapse,zakharov_1974_nonlinear}. In addition, each filament has an apparent double-humped structure, and intense RF electric field also exists on the inner side of these double-humped structures. These double-humps may be caused by intense charge separation~\cite{chen_2016_introduction}.}\miold{Though the wave itself is standing, the envelope is moving, forming a traveling wave packet. The occurrence of the wave packet signifies that the nonlinear effects come into play.}

\begin{figure}[htb]
    \centering
    \includegraphics[width=1\hsize]{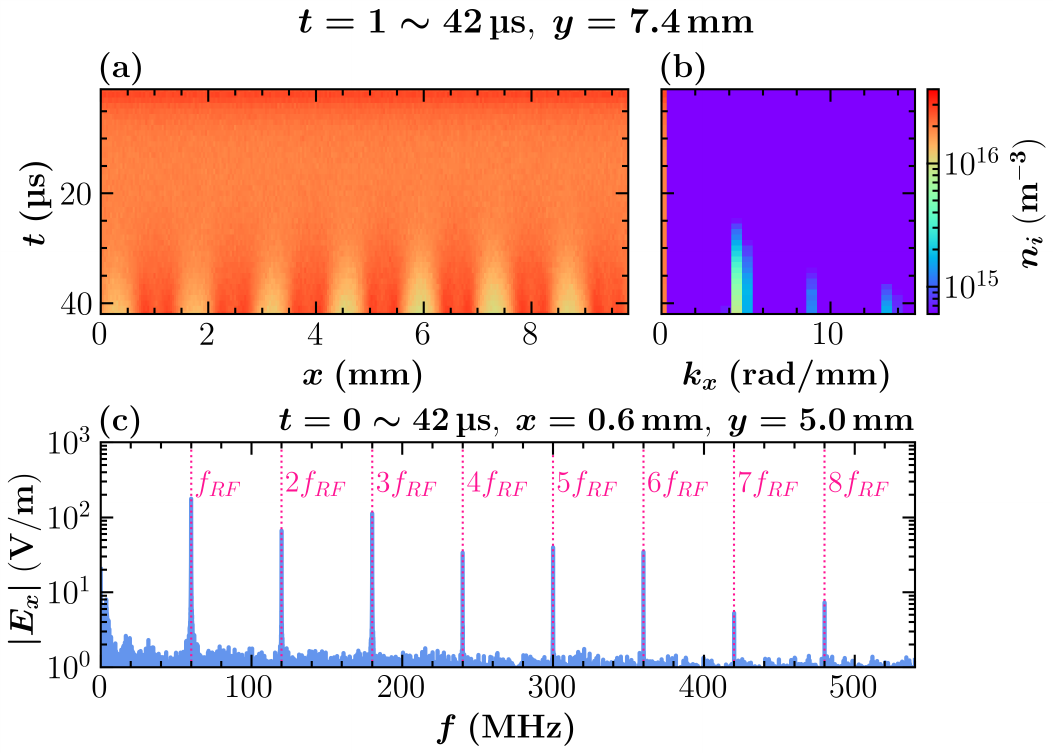}
    \caption{\miold{Wave characteristics}\minew{Fourier spectra} of SWG stage.
    (a) %\minew{Amplifying plasma ripples ($y=L_y/2$) indicate an increasing intensity of the electric field, demonstrating the growth of filamentation instability over time $t = 1 \sim \qty{42}{\us}$.}
    shows the half-height slice of the density varying over time, characterizing the SWG stage.
    And (b) is its wavenumber spectrum evolving over time. Frequency spectrum (c) records the temporal Fourier transform of \miold{radial}\minew{horizontal} electric field $E_x$, measured by a probe at position $(x=\qty{0.6}{mm},\;y=\qty{5.0}{mm})$. The time window of sampling extends from \qtyrange{0}{42}{\us}. The characteristic frequencies are numerically well matched to the RF wave and its higher harmonics.
    }
    \label{Fig3}
\end{figure}

\miold{It should be noted that the observed standing wave differs from the trivial electromagnetic surface wave that propagates radially into the plasma, because the electromagnetic wavelength is much larger than the radius of the calculation domain (see \ref{ApxB}).}
\miold{Instead, it is a type of electrostatic wave that occurs only when the axial magnetic field is applied. To highlight the characteristics of the standing wave, we performed the Fourier transform to the radial electric field $E_x$.}
%\minew{In the SWG stage, the ripple contrast and $E_x$ strength are positively correlated. Amplifying plasma ripples reflect the increasing intensity of RF electric field, demonstrating the growing instability.}
Figure~\ref{Fig3}(a) displays the ion number density evolution within the transformed time window $t = 1 \sim \qty{42}{\us}$ at the slice $y = \qty{7.4}{mm}$, and figure~\ref{Fig3}(b) illustrates the corresponding wave\miold{ }number spectrum \miold{over}\minew{varying with} time. The \miold{radial}\minew{horizontal} uniformity has been maintained until $t = \qty{30}{\us}$, from which moment observable disturbances \miold{surge}\minew{emerge}. Seven filaments form gradually with high regular shapes and nearly equal diameters. The \miold{radial}\minew{transverse} wavelength of these wave packets is\miold{ exact} $\lambda_x = \qty{1.4}{mm}$, i.e. $k_x=2\pi/\lambda_x=\qty{4.49}{\mm^{-1}}$, as depicted in figure~\ref{Fig3}(b). Figure~\ref{Fig3}(c) shows the spectrum of the horizontal component of $\bf{E}$ at position $(x=\qty{0.6}{mm},~y=\qty{5.0}{mm})$\miold{, serving as a typical example. The frequency range spans from 0 MHz to 540 MHz}. Result shows that the characteristic frequencies are exact $f_{\scriptscriptstyle\text{RF}}=\qty{60}{MHz}$ and its higher harmonics.

\miold{Multiple probes positioned at various locations are utilized to capture the temporal oscillations of filamentary plasma parameters during the SWG stage, encompassing}\minew{During SWG stage, multiple parameters at various areas are examined, including} densities, temperatures, velocities, current flux, \miold{electric}\minew{electrostatic} potential, \miold{electrostatic field}\minew{electric fields}, \minew{etc.}\miold{among others.}
Our results show that these parameters share same characteristic frequency spectrum depicted in figure~\ref{Fig3}(c), regardless of probe location. \miold{These spectral profiles underscore the dominance of the wave directly induced by the RF pumping wave (with an RF supply working frequency $f_{\scriptscriptstyle\text{RF}} = \qty{60}{MHz}$) along with its higher harmonics in driving the discharge.}\minew{In these spectra, one can merely observe the characteristic peaks of RF ($f_{\scriptscriptstyle\text{RF}} = \qty{60}{MHz}$) and its harmonics. This spectral pattern highlights the the predominant role of RF pumping.} \miold{This result implies that the significance of }\minew{This suggests that the }cyclotron waves and hybrid waves in this context \miold{is}\minew{are} negligible, since plasma \miold{waves $f_{pe}\approx1.2~\text{GHz}$}\minew{frequencies $f_{pe}=1\sim\qty{3}{GHz}$} and $f_{pi}\approx\qty{4.6}{MHz}$, cyclotron \minew{frequencies}\miold{waves} $f_{ce}=\qty{28}{GHz}$ and $f_{ci}=\qty{0.4}{MHz}$. \miold{Consequently}\minew{To conclude}, the standing wave is solely excited by RF pumping wave.
\minew{This spectral pattern may originate from resonance~\cite{joshi_2018_electron} or modulation~\cite{sun_2022_electron,sun_2022_physical} of RF components. However, in RF discharge experiments, standard detection systems typically filter out RF harmonics, which may mask their existence. %But these RF components possibly hints at unexplored instability mechanisms.
}

% ##### Discussions
%  - 4. Quest: How RF entries TW ? [common CCRF merely manifesting in LW]
%  - 5. Differ to the Diffusion mode.
%  - 6. Explain: Mode Conversion -> NL Wave-Interac. & Bounded Matching
\miold{
In current model, initially uniform plasma of 2D CCRF evolves between radially a pair of periodic boundaries and axially a pair of plat electrode. Both initial and geometric perturbations are excluded. It is thus quite different to the definite conditions of previous simulations~\cite{menati_2019_filamentation,menati_2020_experimentala,menati_2021_variation}. The ion-particle fluid-electron hybrid scheme~\cite{kushner_2009_hybrid} employed in the work~\cite{menati_2019_filamentation} adopted a axially non-symmetric 2D space to model the MDPX device~\cite{thomas_2012_magnetized,thomas_2015_magnetized}, which builds near both metal vertical sides the radial RF electric fields as a strong TW, namely a external perturbation. In contrast, our RF-TW is intrinsic. The two fluid model in works~\cite{menati_2020_experimentala,menati_2021_variation} is performed in a 3D metal box space where over-simplistically omitting RF oscillatory supply, which yields a classical diffusion mode extending into three dimensions. It has produced also concentric filamentary patterns, since the steady-state theory permits the occurrence of higher-order diffusion mode. But it must to be pointed out that its physical basis is entirely different to our results. Our periodic boundaries never satisfies the boundary condition of a diffusion mode on the transverse direction.}

\minew{It should be noted that the periodic configuration has eliminated extrinsic disturbances, such as asymmetric reactor structure in real experiments~\cite{menati_2019_filamentation} and lateral plasma sheaths~\cite{menati_2019_filamentation,menati_2023_formation}. Hence, the occurrence of $E_x$ standing wave should be intrinsic.}

\miold{In this situation}\minew{In conventional consideration}, axial RF pumping should have produced merely a longitudinal wave\miold{ (LW)}\minew{, but here a transverse wave appears}.
%\minew{A question that remains unsolved is how do magnetic fields facilitate the penetration of RF perturbations into the bulk plasma.}
\miold{For this reason}\minew{Therefore}, there should exist \miold{certain}\minew{a} transverse wave \miold{(TW) }mode \minew{that allows}\miold{to which permits the} mode conversion from \miold{RF-LW}\minew{RF longitudinal wave}. The wave mode in which the RF pumping engages is \minew{still }a \miold{theoretical} challenge to \miold{determinate}\minew{determine}, since $f_{\scriptscriptstyle\text{RF}} = 60~\text{MHz}$ differs far from \minew{the }frequency band of \minew{the }usual waves.

\miold{
    It is noteworthy that the radial standing wave generation is highly related to the axial magnetic field. The precise role of the strong magnetic field is complicated and requires a specific study. A rough explanation is that the magnetic field enables the energy transition from the pumping radio-frequency field to some plasma modes, akin to, for example, the generation of helicon waves~\cite{chen_1991_plasma,chen_1997_generalized,arnush_1998_generalized,chen_2015_helicon}.
}

\miold{
    Obvious wave modulation presents commonly in our results. It is usually linked to the occurrence of certain type of MI~\cite{goldman_1984_strong}. The assertion regarding the involvement of nonlinear wave-wave interaction in filamentation is further substantiated through the analysis of wave spectra. Within each frequency spectrum, as exemplified in figure~\ref{Fig3}(c), the characteristic frequencies are integer product of the base-frequency $f_{RF}$. A probable explanation is that the base-frequency $f_{RF}$ wave beating primarily with itself, followed by nonlinear interactions between the RF wave and its higher harmonics, as well as amongst the higher harmonics themselves. The mode matching process selects proper wavenumber and frequency. The temporal-spatial spectra should satisfy the energy conversion conditions, namely $f_1+f_2=f_3$ and ${\bf k}_1+{\bf k}_2={\bf k}_3$, induced from the Manley-Rowe relations, according to the theory of wave non-linearity~\cite{bellan_2006_fundamentals}.
}

\miold{
    Drawing upon a synthesis of simulation outcomes and preceding discourse, a possible qualitative explanation for the filamentation mechanism emerges: the longitudinal RF wave, following mode conversion, infiltrates a transverse component of a specific wave. Facilitated by the mode matching mechanism, this wave manifests a characteristic spatiotemporal frequencies $(f_{RF},\,{\bf k})$. Subsequently, this RF transverse wave engenders modulational instability via nonlinear wave-wave interactions.
}

\minew{
    These results show that the filamentary plasma is an inherently complex system. The self-organization process engages multiple competing mechanisms.
    According to~\cite{menati_2020_experimentala,menati_2021_variation,menati_2023_formation}, an two-fluid diffusion model realized plasma filamentation within a 3D zero-Dirichlet box, despite the exclusion of RF pumping.
    Recent work~\cite{menati_2023_formation} described these filamentary structures as Turing patterns through an activator-inhibitor framework~\cite{satnoianu_2000_turing}. This phenomenological perspective has rather effectively delineated the important role of magnetic field in initiating charge separation and inhomogeneous discharge. However, in contrast to the activation-inhibition-diffusion hypothesis recommended in~\cite{menati_2023_formation}, our simulations do not include the ion-electron recombination effect as inhibition.

    Besides Turing instability, there are other factors potentially related to the pattern formation. As a nonlinear system, RF discharge itself encompasses multiple complexities, such as RF-induced charge separation, RF heating, gas ionization, wave non-linearities, etc.~\cite{lieberman_2005_principles} For instance, modulated ripples are often associated with modulational instabilities~\cite{zakharov_1974_nonlinear,goldman_1984_strong}. Other underlying instabilities, like Kelvin–Helmholtz instabilities in magnetized plasmas and drift instabilities in inhomogeneous plasmas~\cite{treumann_1997_advanced}, may also contributed to the growth of plasma ripples.

    Furthermore, simulations~\cite{menati_2020_experimentala,menati_2023_formation} suggest that the filamentation phenomenon is not solely attributable to RF discharge and exhibits broader universality. Perhaps, filamentation may occur in other working conditions, such as direct-current discharge~\cite{jin_2022_particleincell}. This is a topic warranting further exploration.
}

\subsection{Pattern Evolving Stage}\label{Sec3B}
% ** B. PE stage
%  # Results
%  - 1. Characterizing PE stage
%  - 2. Stating mechanics
%  - 3. Summarizing dynamics
%  # Discussions
%  - 4. Outlooks

In the pattern evolving \minew{(PE) }stage, \miold{the entire system undergoes turbulence since the end of the SWG stage. }\minew{the}\miold{Bulk} plasma transits to non-equilibrium dynamical system. The filaments can flow \miold{chaotically}\minew{erratically}, merge in pairs or even surge suddenly from depletion region, suggesting their highly nonlinear nature. Regularity is disrupted as the number and positioning of filaments \miold{cease to}\minew{no longer} remain constant. Yet filaments maintain nearly equal diameters\minew{ and amplitudes of the same order.}\miold{, and the saturation density limitation persists.}

% ##### Results
\miold{While the intricate trajectory of flow remains still unpredictable, an overarching trend emerges, driving the compression of filaments to coalesce into a system characterized by fewer number yet more pronounced amplitude of filaments. In our simulation, the equilibrated number of filaments hovers around four, yet a real steady-state never reached despite being quasi-steady in time-span of 0.1 ms.}\minew{Although the filaments flow chaotically, pattern dynamics tend to favor the coalescence of filaments into a smaller number yet with stronger magnitude. Our simulation stabilizes at around 4 filaments. Throughout the PE stage, filamentary plasma exhibits quasi-steady oscillations over a \qty{0.1}{\us} cycle but never attains a true steady-state.}

\begin{figure*}[ht]
    \centering
    \includegraphics[width=0.85\textwidth]{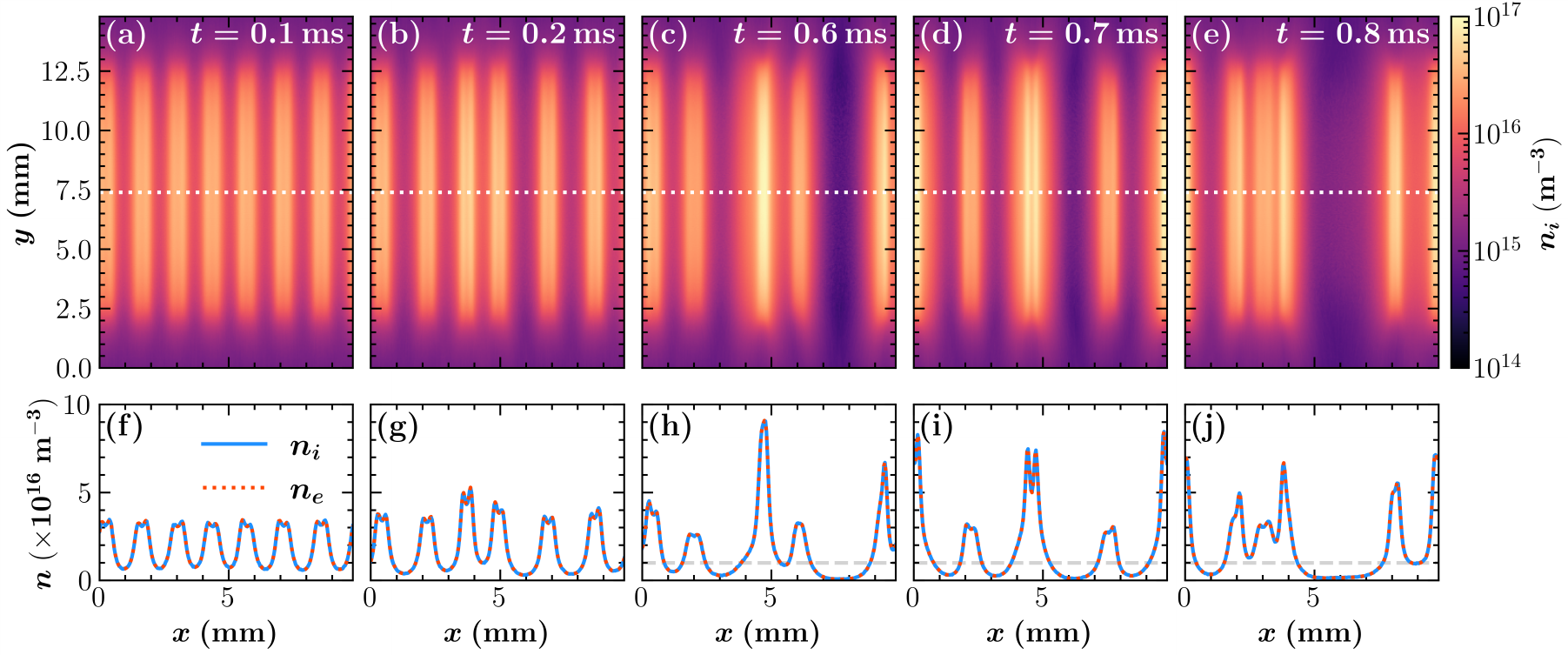}
    \caption{Pattern evolving stage.
    The $\text{Ar}^+$ 2D profiles present the plasma distribution at distinct moments: $t$ = (a)~\qty{100}{ms}, (b)~\qty{200}{ms}, (c)~\qty{600}{ms}, (d)~\qty{700}{ms}, (e)~\qty{800}{ms}. The white dot-dashed lines highlight the half-height altitude $y=\qty{7.4}{mm}$. Blue and red curves indicating respectively $\text{Ar}^+$ and $\text{e}^-$ depicted in (f--j) are their corresponding 1D profiles along the half-height slice. The grey dashed line in (h--j) is the density level of \qty{1e16}{m^{-3}}, as a criterion to distinguish regions of filament or depletion.}
    \label{Fig4}
\end{figure*}

\minew{
    Now, we focus on the time-averaged effects of RF dynamics and use operator $\langle~\rangle$ to denote the time-averaging over multiple RF-periods. We define for any quantity $\clubsuit$,
    \begin{equation*}
    \braket{\clubsuit}({\bf x},t)=\tau^{-1}\int^{t+\tau/2}_{t-\tau/2}{\clubsuit}({\bf x},t'){\rm d}t',
    \end{equation*}
    with duration $\tau = 6 T_{RF} = \qty{0.1}{\us}$.
}

\begin{figure}[htb]
    \centering
    \includegraphics[width=1\hsize]{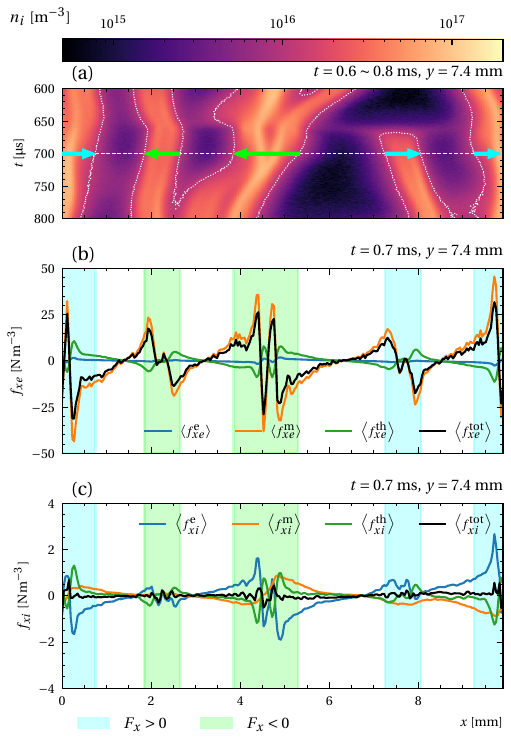}
    \caption{
        Fluid dynamics of PE stage. The\miold{se} half-height slices constitute (a) the $\text{Ar}^+$ density profile varying over time $t = 0.6\sim\qty{0.8}{\ms}$, which displays the hydro-kinematics of PE stage. White dot-dashed contours close the regions of ion density exceeding \qty{1E16}{m^{-3}} where are viewed as plasma filaments.
        In (a), the rightward cyan or leftward lime arrow attached upon each filament indicates the direction of total force $F_x$ acting on whole filament at the moment \qty{700}{\us}.
        (b) \miold{demonstrates}\minew{ and (c) demonstrate} the local volume force\minew{s} along the waistline $y=L_y/2$ at \qty{700}{\us}, including the \minew{time-averaged }resultant force\minew{s} (\miold{${f}^\text{tot}_{ex}$}\minew{$\braket{{f}^\text{tot}_{ex}}$ and $\braket{{f}^\text{tot}_{ix}}$}, black curve\minew{s}) and \miold{two}\minew{three kinds of} dominant forces, that is, \miold{the time-averaged Lorentz force ($\braket{{f}^\text{em}_{ex}}$, magenta curve)}\minew{the time-averaged electrical force ($\braket{{f}^\text{e}_{x}}$, blue curve), the time-averaged magnetic force ($\braket{{f}^\text{m}_{x}}$, orange curve),} and the time-averaged thermal pressure force (\miold{$\braket{{f}^\text{th}_{ex}}$, azure}\minew{$\braket{{f}^\text{th}_{x}}$, green} curve) exerted upon electrons\minew{ and ions}.
        In \miold{(b)}\minew{both (b) and (c)}, lime or cyan patch spanning entire region of each filament, means that the total force acting on the whole filament is pointing to left-hand-side or right-hand-side, respectively.
    }
    \label{Fig5}
\end{figure}

At the microsecond scale, \miold{the bulk plasma distribution undergoes minimal changes, meaning that the temporal derivative of density $\partial_t n_s$ approaches zero.}\minew{the time-averaged density $\braket{n_s}$ remains basically unchanged.}
\miold{In the local conservation equation of charged species $s\in \{\text{Ar}^+,\,\text{e}^- \}$
\begin{equation}
  \partial_t n_s + \nabla \bm{\cdot} n_s {\bf U}_s = s_s ~, \label{Eq1}
\end{equation}
the electron source $s_e$ and ion source $s_i$ are equal within bulk plasma. Moreover, source term $s_s$ is equal to ionization rate $n_e\nu^\text{ioniz}$ where $\nu^\text{ioniz}$ is local ionization frequency. Computational results reveal that the ionization rate $n_e\nu^\text{ioniz}$ and the flux divergence $\nabla \bm{\cdot} n_s {\bf U}_s$ are in comparable strength.}
\miold{This result indicates that the plasma is characterized by the relatively weak transport and the high local ionization efficiency, resulting in a high-density region.}
\minew{However, i}\miold{I}t can be observed that new plasma filament\minew{s} occasionally emerges from the depletion region\minew{ (e.g., in figure~\ref{Fig5}(a))}, owing to \miold{the local high temperature}\minew{high local ionization rate} and \minew{magnetic-confined horizontal transport}\miold{a weakened initial diffusion}.

Figure~\ref{Fig4} shows the plasma \minew{density}\miold{distribution} snapshots in the PE stage. Two groups of sub-figures~\ref{Fig4}(a--b) and sub-figures~\ref{Fig4}(c--e) present separately the transition from SWG to PE stage and a highly chaotic scenario. Their corresponding density curves are plotted in sub-figures~\ref{Fig4}(f--j) respectively. At \qty{0.1}{ms}, seven filaments are regularly arranged (figure~\ref{Fig4}(a)) and their density peak are of equal height (figure~\ref{Fig4}(f)); But at \qty{0.2}{ms}, only six irregular filaments distribute erratically (figure~\ref{Fig4}(b)), their density values start to differ as well (figure~\ref{Fig4}(g)). Such transition is attributed to the motion and pairwise merging of filaments. Having experienced a characteristic evolution time about 0.1 millisecond, a pair of filaments meld into single one, the charged particle numbers stack up, a brighter filament is left, as clearly drawn in figures~\ref{Fig4}(f--g). Next, the scenario projected onto figures~\ref{Fig4}(c--e) shows a representative instance of chaotic hydro-kinematics, depicted through the evolution of the ion distribution during time slot $t = 0.6 \sim\qty{0.8}{ms}$, capturing various phenomena including chaotic flowing, pairwise merging, sudden emerging and wave saturating. At these moments, some packets flow together and locate close enough to dynamically view them as a whole. \miold{Density level $n_c = \qty{1e16}{m^{-3}}$ is thus introduced as a quantitative yardstick to divide the regions of filament and depletion. For instance, a connected area satisfying $n_i > n_c$ would be seen as a whole filament since it has concentrated majority of mass in the vicinity.}

\minew{The plasma densities of the filament and the depletion region differ greatly. Typically, in depletion regions, the plasma density is below $\qty{1e16}{m^{-3}}=:{n_c}$, thus we define this density level as a criterion for demarcating the filament and depletion regions. A contiguous region with $n_i > n_c$ is regarded as a filament. %This region contains the majority of the mass nearby.
}

\miold{To start, sub-figure~\ref{Fig4}(c) reveals five filaments, already exhibiting irregular positioning and varying magnitudes. Subsequently, as detailed in the dynamically-evolving scenario presented in figure~\ref{Fig5}(a), the plasma mass gravitates towards the three filaments highlighted in figure~\ref{Fig4}(e) finally amidst the complete chaos. A quintessential instance of ``sudden emerging" manifests in figure~\ref{Fig4}(d), notably observed with the second filament from the right suddenly emerging from the depletion region at around 0.66~ms (See figure~\ref{Fig5}(a)). This occurrence unfolds abruptly and progresses swiftly, devoid of any discernible precursor indications at previous moment depicted in figure~\ref{Fig4}(c).}\minew{
    Figure~\ref{Fig4}(c) initially captures five filaments with irregular spatial positioning and distinct peak-density. During fully nonlinear development (figure~\ref{Fig5}(a)), the plasma aggregates towards three brightest filaments in ~\ref{Fig4}(e). A characteristic ``sudden emerging'' event occurs in figure~\ref{Fig4}(d), where the second rightmost filament abruptly materializes from the depletion zone at 0.66 ms (figure~\ref{Fig5}(a)). This transient phenomenon initiates without observable precursors in prior frames (figure~\ref{Fig4}(c)), exhibiting sudden onset and rapid progression.
}

\minew{To analyze the non-equilibrium pattern dynamics, we commence with a two-fluid model of magnetized cold plasma. }The complete fluid momentum balance equation for charged species \minew{in cold plasma is composed of six}\miold{consists of seven} mechanical terms, namely, \miold{inertia $\partial_t m_s n_s {\bf U}_s$, convection ${\bf f}^\text{conv}_s$, stresses (both thermal pressure ${\bf f}^\text{th}_s$ and viscosity ${\bf f}^\text{visc}_s$), collisions ${\bf f}^\text{coll}_s$, gravity ${\bf f}^\text{gr}_s$ and electromagnetism ${\bf f}^\text{em}_s$, as referenced in the textbook~\cite{nicholson_1983_introduction} and literature~\cite{yang_2017_energy,smolyakov_2017_fluid}.}
\minew{
\begin{itemize}
\item[(a)]inertia $\partial_t m_s n_s {\bf U}_s$,
\item[(b)]convection ${\bf f}^\text{conv}_s = - ({\bf U}_s \bm{\cdot} \nabla) m_s n_s {\bf U}_s$,
\item[(c)]thermal pressure ${\bf f}^\text{th}_s = - k_B \nabla \bm{\cdot} n_s {\bf\hat{T}}_s$,
\item[(d)]gyro-viscosity ${\bf f}^\text{gyro}_s = - \nabla \bm{\cdot} \bm{\hat{\Pi}}^\text{gyro}_s$,
\item[(e)]neutral collision ${\bf f}^\text{n}_s = - m_s n_s \nu^{s\text{-}n}{\bf U}_{s}$,
\item[(f)]electromagnetism ${\bf f}^\text{em}_s = q_s n_s ({\bf E} + {\bf U}_s \times {\bf B})$.
\end{itemize}
These are based on what's described in the textbooks~\cite{nicholson_1983_introduction,treumann_1997_advanced,chen_2016_introduction} and literature~\cite{yang_2017_energy,smolyakov_2017_fluid,xu_2023_rotating}. Here, the gyro-viscosity $\bm{\hat{\Pi}}^\text{gyro}_s$~\cite{chew_1956_boltzmann} contributes the major component of the shear stress, with due consideration of the Finite Larmor Radius effect~\cite{cheng_1999_kineticfluid,smolyakov_2017_fluid}. The formalism for numerical calculation of gyroviscosity tensor $\bm{\hat{\Pi}}^\text{gyro}_s$ ultimately adopted the widely used formulation outlined respectively in \cite{smolyakov_2017_fluid}, consistent with that detailed in \cite{xu_2023_rotating}. In addition, the major type of collision between the cold plasmas and static neutral background is the neutral collision term~\cite{chen_2016_introduction} under the neutral collision frequency $\nu^\text{n-s}$.
}The momentum equation is thus
\begin{equation}
  \partial_t m_s n_s {\bf U}_s = {\bf f}^\text{conv}_s + {\bf f}^\text{th}_s + {\bf f}^\text{\minew{gyro}\miold{visc}}_s + {\bf f}^\text{\minew{n}\miold{coll}}_s \miold{+ {\bf f}^\text{gr}_s} + {\bf f}^\text{em}_s \miold{,}\minew{.} \label{Eq2}
\end{equation}
\miold{where convection term is ${\bf f}^\text{conv}_s = - ({\bf U}_s \bm{\cdot} \nabla) m_s n_s {\bf U}_s$, thermal pressure force is ${\bf f}^\text{th}_s = - k_B \nabla \bm{\cdot} n_s {\bf\hat{T}}_s$, viscous force is ${\bf f}^\text{visc}_s = - \nabla \bm{\cdot} \bm{\hat{\Pi}}_s$, \miold{gravitation is ${\bf f}^\text{gr}_s = m_s {\bf g}$} and Lorentz force is ${\bf f}^\text{em}_s = q_s n_s ({\bf E} + {\bf U}_s \times {\bf B})$.}

\miold{This fluid motion equation is introduced to discuss the hydrodynamics involving in non-equilibrium filaments. However, this universal formulation needs certain approximations for convenient determination. }%\minew{This fluid motion equation is introduced with the aim of discussing the hydrodynamics involved in non-equilibrium filaments. Nevertheless, for the sake of easy consideration, this universal formulation requires certain simplifications.}
\miold{In the present scenario, gravitation never functions. }\miold{The shear stress force predominantly originates from its gyroviscous component $\bm{\hat{\Pi}}^\text{gyro}_s$~\cite{chew_1956_boltzmann}, i.e. ${\bf f}^\text{visc}_s \approx {\bf f}^\text{gyro}_s = - \nabla \bm{\cdot} \bm{\hat{\Pi}}^\text{gyro}_s$, with due consideration of the FLR effect~\cite{cheng_1999_kineticfluid,smolyakov_2017_fluid}. The formalism for numerical calculation of gyroviscosity tensor $\bm{\hat{\Pi}}^\text{gyro}_s$ ultimately adopted the widely used formulation outlined respectively in \cite{smolyakov_2017_fluid}, consistent with that detailed in \cite{xu_2023_rotating}. Notably, the majority of collisions in partly-ionized plasmas stem from the neutral collisions term, that is ${\bf f}^\text{coll}_s \approx {\bf f}^\text{n}_s = m_s n_s \nu^\text{n-s} {\bf U}_s$ with neutral collision frequency $\nu^\text{n-s}$~\cite{chen_2016_introduction}.}

\miold{
    Slow dynamics of the patterns only concerns the time-averaged effects of RF plasma oscillations. We use angle brackets $\langle~\rangle$ to denote the operator for time-averaging physical quantities over duration exceeding multiple RF periods, for instance, $\braket{\Box}=\int^{\tau}_{0}{\Box}\rm{d}t$ as $\tau = 6 T^{-1}_{RF} = \qty{0.1}{\us}$.
}

The time-averaged convection $\braket{{\bf f}^\text{conv}_s}$ is estimated by the ponderomotive force ${\bf f}^\text{pm}_s$ i.e. $\braket{{\bf f}^\text{conv}_s} \approx {\bf f}^\text{pm}_s$\minew{.}\miold{, as substituting the plasma dispersion relation into the expression for the convective force $\braket{{\bf f}^\text{conv}_s} \approx - m_s n_s \nabla \langle{\bf U}^2_s\rangle$ yields the expression for the electrostatic PM force ${\bf f}^\text{pm}_s = - {\omega^2_{ps}} \nabla \langle {\bf E}{\cdot}{\partial_{\omega}\bm{\hat{\epsilon}}}{\cdot}{\bf E} \rangle$~\cite{maxwell_2010_treatise,lundin_2006_ponderomotive}. Here, the plasma dielectric tensor $\bm{\hat{\epsilon}} = \epsilon_0\big{[}{\bf b}\otimes{\bf b} + ({\bf\hat{I}}-{\bf b}\otimes{\bf b})\big{/}(1-\omega^2_{cs}/\omega^2)^2\big{]}$ is function of $\omega$ with the unit matrix $\bf{\hat{I}}$, the magnetic field direction unit vector ${\bf b}$ and the cyclotron frequency $\omega_{cs}$, and the plasma frequency $\omega_{ps}=\sqrt{\frac{n_s q^2_s}{\epsilon_0 m_s}}$ is circular frequency of Langmuir wave.}
\minew{
    This approximation arises by substituting the dispersion relation into the expression for the convective force
    \begin{equation*}
        \braket{{\bf f}^\text{conv}_s} = - m_s \braket{n_s ({\bf U}_s\bdot\nabla) {\bf U}_s} \approx - \frac{1}{2} m_s \braket{n_s} \nabla \braket{{\bf U}^2_s},
    \end{equation*}
    which ultimately yields the electrostatic ponderomotive force~\cite{maxwell_2010_treatise,lundin_2006_ponderomotive,dasgupta_2003_kinetic}
    \begin{equation*}
    {\bf f}^\text{pm}_s = - \frac{\braket{n_s} q^2_{s}}{4 m_{s}} \nabla \Braket{
        \frac{E_\perp^2}{\omega^2-\omega^2_{cs}} + \frac{E_{\scriptscriptstyle{\parallel}}^2}{\omega^2}
    },
    \end{equation*}
    where $\omega_{cs}$ is the cyclotron frequency, $\omega$ is the wave frequency, and the electric field components $E_\perp$ and $E_{\scriptscriptstyle{\parallel}}$ are perpendicular and parallel to the magnetic field ${\bf B}$, respectively.
}
\miold{The time-derivative term of the ponderomotive force~\cite{washimi_1976_ponderomotive,tskhakaya_1981_nonstationary,kentwell_1987_timedependent} can be negligible in slowly varying RF-averaged waveforms.}\minew{Term $\braket{{\bf E}^2}$ is slowly varying in the PE stage, thus time-derivative term of the ponderomotive force~\cite{washimi_1976_ponderomotive,tskhakaya_1981_nonstationary,kentwell_1987_timedependent} is neglected.} Contributions from anisotropic temperature~\cite{ghildyal_1998_ponderomotive} and non-local properties~\cite{smolyakov_2007_nonlocal,iwata_2013_nonlocal} are \minew{also }disregarded.

The time-averaged Lorentz force \miold{$\langle {{\bf f}^\text{em}_s}\rangle = q_s \langle{n_s}\rangle \langle{\bf E} + {\bf U}_s \times {\bf B}\rangle$}\minew{$\braket{{\bf f}^\text{em}_s} = \braket{{\bf f}^\text{e}_s}+\braket{{\bf f}^\text{m}_s} = q_s \braket{n_s{\bf E}} + q_s \braket{n_s{\bf U}_s} \times {\bf B}$} sees the time-averaged fields $\langle\minew{n_s}{\bf E}\rangle$ and $\langle\minew{n_s}{\bf U}_s\rangle$ where the high frequency oscillations are filtered out by time-averaging. \minew{Via}\miold{Using} the same method, we calculate the time-averaged values of thermal pressure force $\langle{\bf f}^\text{th}_s\rangle = - k_B \nabla \bm{\cdot} \langle n_s {\bf\hat{T}}_s\rangle$, gyroviscous force $\langle{\bf f}^\text{gyro}_s\rangle = - \nabla \bm{\cdot} \langle\bm{\hat{\Pi}}_s\rangle$, and neutral collisional friction $\langle{\bf f}^\text{n}_s\rangle = m_s \langle n_s {\bf U}_s\rangle \nu^\text{n-s}$. The time-averaged right-hand side of equation~(\ref{Eq2}) is \miold{therefore}\minew{finally}
\begin{equation}
\minew{\langle}{\bf f}^\text{tot}_s\minew{\rangle} = {\bf f}^\text{pm}_s + \langle{\bf f}^\text{th}_s\rangle + \langle{\bf f}^\text{gyro}_s\rangle + \langle{\bf f}^\text{n}_s\rangle + \langle{\bf f}^\text{em}_s\rangle , \label{Eq3}
\end{equation}
where $\minew{\langle}{\bf f}^\text{tot}_s\minew{\rangle}$ represents the \miold{total}\minew{effective resultant} force\minew{, averaged over duration $\tau=\qty{0.1}{\us}$, }experienced by \minew{local fluid element.}\miold{the fluid on a microsecond time scale.}

\miold{
    A numerical evaluation is conducted to compare the time-averaged contributions of five forces in equation~(\ref{Eq3}), delving into the theoretically pertinent forces governing plasma manipulation along the radial axis. Our results indicate that the pattern dynamics in PE stage is primarily governed by time-averaged Lorentz force $\langle{\bf f}^\text{em}_s\rangle$ and time-averaged thermal pressure force $\langle{\bf f}^\text{th}_s\rangle$.
}
%A numerical evaluation is conducted to compare the time-averaged contributions of five forces in equation~(\ref{Eq3})\miold{, delving into the theoretically pertinent forces governing plasma manipulation along the \miold{radial}\minew{horizontal} axis}. Our results indicate that the pattern dynamics\minew{, of both ions and electrons,} in PE stage is primarily governed by time-averaged Lorentz force $\langle{\bf f}^\text{em}_s\rangle$ and time-averaged thermal pressure force $\langle{\bf f}^\text{th}_s\rangle$.

\miold{Figure~\ref{Fig5} explores the relationship between the motion of filaments and the forces acting upon them. }Figure~\ref{Fig5}(a) presents the patterns flowing over time $t = 0.6 \sim \qty{0.8}{ms}$ and \miold{the total force direction}\minew{the direction of the total force acting on each filament at $t=\qty{0.7}{ms}$}. \miold{The spatiotemporal evolution graph displays the half-height slice of the ion density varying over time $t = 0.6 \sim \qty{0.8}{ms}$ (in the PE stage). }This scenario has been projected its moments $t$ = 0.6, 0.7, 0.8~ms onto figures~\ref{Fig4}(c--e). \miold{It presents a good sample delving into the chaotic hydro-kinematics of PE stage. }Various phenomena including \minew{(i) }chaotic flowing, \minew{(ii) }pairwise merging, \minew{(iii) }sudden emerging and \minew{(iv) }wave saturating are clearly captured here. White dot-dashed curves enclose the regions of ion density exceeding $\minew{n_c =}$~\qty{1e16}{{m}^{-3}} where are tagged as filament areas for the force calculation.

\minew{
The total force in the $x$-total, denoted as $F_x$, acting on each filament, is defined as the spatial integral of the horizontal component of the local total force $\braket{{\bf f}^\text{tot}_{ix}}+\braket{{\bf f}^\text{tot}_{ex}}$ acting on each filamentary region
\begin{equation*}
F_x =\int_\text{filament region}\braket{{f}^\text{tot}_{ex}+{f}^\text{tot}_{ix}}{\rmd}x{\rmd}y.
\end{equation*}
If $F_x > 0$, the total force $F_x$ is directed to the right; if $F_x < 0$, the total force $F_x$ is directed to the left.
}

\miold{Area total force $F_x$ is defined by the filament area integral of the radial component of local total force ${\bf f}^\text{tot}_{e}$ acting on fluid electrons, namely, $F_x =\int_\text{filament}{f}^\text{tot}_{ex}{\rm d}x{\rm d}y$. If $F_x > 0$, it indicates that the total force is directed to the right, whereas if $F_x < 0$, it indicates that the force is directed to the left.} The direction of \minew{$x$-}resultant force acting on whole filament \minew{$F_x$ }at moment $t = \qty{0.7}{ms}$ is indicated by the colored arrows overlaid on the \miold{heat map }background\minew{ in figure~\ref{Fig5}(a)}. \miold{As illustrated in figure~\ref{Fig5}(a),}\minew{Here,} the acceleration \miold{direction of filament areas is exactly coherent to the direction of the total force.}\minew{of each filament is in the same direction as the calculated resultant force $F_x$ acting on it.}
Figure~\ref{Fig5}(b) displays the calculated result details of effective volume forces acting along the half-height level $y=L_y/2$ at the moment $t = \qty{0.7}{ms}$. It includes the local total force ($\minew{\langle}f^\text{tot}_{ex}\minew{\rangle}$, shown as the black curve) and \miold{two}\minew{three} dominant forces: \minew{the $\braket{{f}^\text{e}_{ex}}$, $\braket{{f}^\text{m}_{ex}}$ and $\braket{{f}^\text{th}_{ex}}$.}\miold{the time-averaged Lorentz force ($\langle {f}^\text{em}_{ex}\rangle$, represented by the magenta curve) and the time-averaged thermal pressure force ($\braket{ {f}^\text{th}_{ex}}$, shown as the azure curve) acting upon electrons.} In \minew{addition,}\miold{Besides,} a lime or cyan patch spanning the entire region of each filament indicates that the area total force $F_x$ acting on the whole filament is directed towards the left-hand side or right-hand side, respectively. \minew{Figure~\ref{Fig5}(c) has the same format as figure~\ref{Fig5}(b), but it shows the force conditions of ions.}

\miold{
   In the simulation results, at $t$ = \qty{0.7}{ms}, the plasma parameters along the half-height line are quasi-stationary, exhibiting negligible changes within \qty{\pm1}{\us}. Employing the average intensities of the five forces in the vicinity of $t$ = \qty{0.7}{ms}, spanning 30 RF cycles (equivalent to 0.5~\textmu s), facilitates a straightforward comparison of the contributions of these five forces to the local plasma energy change. The results reveal that these five forces are not balanced within the bulk plasma. In other words, the inertia term is always non-zero, which argues that the plasma of PE stage is far from a force-balanced system. The time-averaged thermal pressure force $\langle{\bf f}^\text{th}_s\rangle$ and the time-averaged Lorentz force $\langle{\bf f}^\text{em}_s\rangle$ significantly outweigh the other three forces (by three orders of intensity at least). Therefore, these two forces is sufficient to approximate the resultant force acting on the plasma. The thermal pressure force $\langle{\bf f}^\text{th}_s\rangle$ and the Lorentz force $\langle{\bf f}^\text{em}_s\rangle$ are oriented in the inverse direction locally, which eventually drives the particles towards higher-density areas. At larger scale, the sum of forces acting on each particle within the plasma filament is used to represent the resultant force acting on the entire filament, propelling the motion of the whole filament.
}

\minew{
    Around the simulation time $t = \qty{0.7}{ms}$, within a time interval of \qty{\pm1}{\us}, the plasma near the half-height line $y=L_y/2$ are regarded quasi-steady. However, on the scale of 100 microseconds, one can observe that the filamentary or rod-shaped flow clusters exhibit obvious collective speed changes. Thus, we calculate the force acting on the system at this moment and compare it with the simulation results. The force characteristics of electrons are summarized as follows: (i) $\braket{{\bf f}^\text{tot}_e}$ is not balanced and contributes the largest share to $F_x$; (ii) More specifically, $\braket{{\bf f}^\text{m}_e}$ contributes the largest share to $\braket{{\bf f}^\text{tot}_e}$, followed by $\braket{{\bf f}^\text{th}_e}$, and then comes $\braket{{\bf f}^\text{e}_e}$; (iii) ${\bf f}^\text{pm}_e$, $\braket{{\bf f}^\text{gyro}_e}$ and $\braket{{\bf f}^\text{n}_e}$ are at least three orders of magnitude smaller than $\braket{{\bf f}^\text{th}_e}$, therefore neglected; (iv) The trend generally shows that $\braket{{\bf f}^\text{th}_e}$ and $\braket{{\bf f}^\text{e}_e}$ jointly resist $\braket{{\bf f}^\text{m}_e}$; (v) $\braket{{\bf f}^\text{tot}_e}$ transports electrons along the density gradient $\partial_x n_p$. %These findings suggest that the effective $z$-flux $\braket{{n_e}{U_{ez}}}$ is at the core position in the pattern dynamics of filamentation.
}

\minew{
    Then, we compared the force profiles of ions and electrons and found that: (i) $\braket{{\bf f}^\text{em}_i}=\braket{{\bf f}^\text{e}_i}+\braket{{\bf f}^\text{m}_i}$ and $\langle{\bf f}^\text{th}_i\rangle$ are nearly in balance; (ii) The maximum magnitude of $\braket{{\bf f}^\text{tot}_i}$ is $\leq 5\%$ of that of $\braket{{\bf f}^\text{tot}_e}$; (iii) ${\bf f}^\text{pm}_i$, $\braket{{\bf f}^\text{gyro}_i}$ and $\braket{{\bf f}^\text{n}_i}$ are also very small and thus neglected; (iv) The curve shape of $\braket{{\bf f}^\text{th}_i}$ are highly similar to $\braket{{\bf f}^\text{th}_e}$; (v) The directions of $\braket{{\bf f}^\text{e}_i}$ and $\braket{{\bf f}^\text{m}_i}$ are opposite to those forces acting on electrons, respectively. %Thus, when analyzing the slow dynamics of the filamentary patterns, it suffices to take into account the force fields $\langle{\bf f}^\text{em}_x\rangle$ and $\langle{\bf f}^\text{th}_x\rangle$ that act on the electrons.
}

\minew{
    Remarkably, $\braket{{\bf f}^\text{m}_e}$ and $\braket{{\bf f}^\text{m}_i}$ reveal that $\braket{E_{x}}$ generate strong $\braket{E}\times{B}$ shear flows twist around each plasma filament. The $\braket{U_{ez}}$ in the depletion region is much larger than $\braket{U_{ex}}$. This flow may be an important factor in the formation of filamentary structures, for example, it could fragment plasma into small spots. This ${E}\times{B}$ shear flow is a common feature in plasma filamentation, and its existence was already implied in previous experimental observations~\cite{schwabe_2011_pattern,williams_2022_experimental}.
}

\minew{
    These results reveal that the total force $\braket{{\bf f}^\text{tot}_e + {\bf f}^\text{tot}_i}$ are not balanced within the bulk plasma. In other words, the inertia term is always non-zero, which argues that the plasma of PE stage is far from a force-balanced system. The time-averaged thermal pressure force $\langle{\bf f}^\text{th}_s\rangle$ and the time-averaged Lorentz force $\langle{\bf f}^\text{em}_s\rangle$ significantly larger than the other three forces (by three orders of intensity at least). Therefore, these two forces are sufficient to approximate the resultant force acting on the plasma. The thermal pressure force $\langle{\bf f}^\text{th}_s\rangle$ and the Lorentz force $\langle{\bf f}^\text{em}_s\rangle$ are oriented in the inverse direction locally, which eventually drives the particles towards higher-density areas. At a larger scale, the total force $F_x$ acting on each filament propels its integral motion in a manner similar to rigid body.

    % Finally, we use figure~\ref{Fig8}, a force diagram of the filamentary pattern, to summarize our key findings.
    % \input{Fig8}
}

\miold{
    Despite the relative weakness of ponderomotive force ${\bf f}^\text{pm}_s$, the time-average term $\frac{\epsilon_{0}}{2} \langle E_x^2 \rangle$, representing radiation pressure, exhibits a steep gradient within the plasma bulk and also the electric field envelope appears akin to a solution of the nonlinear Schrödinger equation found within the set of Zakharov equations~\cite{zakharov_1972_collapse,zakharov_2013_nonlinear,gelash_2014_superregular}. Generally, in steady-state plasmas described by Zakharov equations, the gradients of thermal and radiation pressures oppose each other~\cite{nicholson_1983_introduction,treumann_1997_advanced}. This implies that the PM effect tends exert pressure on the lower-density regions, while the thermal pressure force acts to smooth out density gradients. However, the radiation pressure does not match this relationship in our results; its spatial fluctuations align closely with the radial thermal pressure $k_B n_e T_{ex}$ throughout the domain. Hence, attempts to adapt the Zakharov equations to describe the filamentation are certain to fail.
}
\section{Conclusions}\label{Sec4}

In this study, we utilized 2D PIC/MCC simulations to investigate the filamentation process of\miold{ CCRF} magnetized \minew{RF }plasma and analyze the associated pattern dynamics.

%\minew{The current work is devoted to revealing the physical commonalities of this phenomenon. To achieve this, we simulated the filamentation on different models above all and we have examined the representativeness of the model of the periodic parallel electrodes. On this basis, we presented and discussed the simulation results of the periodic model in detail.}

We observed \minew{two dynamic regimes}\miold{distinct stages} in the \minew{complete filamentation }process\miold{, delineating the Standing Wave Growing stage and the subsequent Pattern Evolving stage, with the moment of instability saturation marking a critical transition point}. Initially, longitudinal filaments emerge from a uniform background\minew{ and amplify until saturation. }\miold{, while in the Pattern Evolving stage}\minew{Thereafter}, filaments can flow, grow, merge, and emerge suddenly from depletion regions.

\miold{Regarding wave characteristics, our results indicate that the non-linear oscillations, of such as plasma and electrostatic field, exhibit waveform and spectrum properties of modulational instability in nonlinear wave theory. Self-modulation provides these characteristic spectra consisted of integer multiples of the base spatiotemporal frequencies. Though the exact wave mode remains unidentified, these findings partially reveal the instability mechanism.}

\minew{
    Our results show that the plasma ripples and RF electrostatic standing waves are modulated. Moreover, each filament equips a double-humped peak; intense RF electric field occurs insides these structures. The spectrum demonstrates that the oscillations are generally of multiple-RF frequencies.
}

\miold{Fluid mechanics}\minew{Analysis of different terms of fluid momentum equation} reveals that \miold{time-integrated}\minew{time-averaged} forces such as Lorentz and thermal pressure dominate during the Pattern Evolving stage. The competition between \miold{these two terms}\minew{electrical, magnetic and thermal pressure forces} governs the motions of filaments.\miold{ Additionally, results suggest that perturbed local ionization rate outweighs magnetic-confined radial diffusion, thereby facilitating sudden emergence of filament from a depletion region.}
\minew{Through RF-cycle averaging, our analysis demonstrated that electrons and ions are governed by the magnetic force and electric force respectively. The time-averaged magnetic force drives electrons to accumulate at plasma density maxima, while time-averaged electric force pushes ions into the same regions, jointly molding the filaments.
}

These novel \minew{insights}\miold{clews} offer valuable guidance for future investigations into pattern dynamics and essential reference for effective manipulations of industrial plasmas.
\minew{Our findings can provide the knowledge regarding the background magnetized plasma for dynamic control of the dust therein.}
%\miold{The validation of wave non-linearity}\minew{Our results on the RF electric wave structure} is waiting for future experimental \miold{measurement}\minew{validation}.
\miold{The validation of wave non-linearity is waiting for future experimental measurement.
Elucidating the intricacies of magnetic instability constitutes the forthcoming focal point of our future theoretical endeavors.}
\miold{Moreover, f}\minew{F}uture research should include 3D kinetic simulations to explore filament micro-structures and extend investigations to electromagnetic cases.

% Acknowledgements
\ack
We thank Prof. L. Xu~\orcidlink{0000-0002-8729-1984}, doctoral students J.-H. Chen and B.-M. Jin~\orcidlink{0000-0002-9465-4480}, for constructive sharing of thoughts. This work is supported by the National Natural Science Foundation of China (Grant Nos. 12305223, 12175322) and the National Natural Science Foundation of Guangdong Province (Grant No.2023A1515010762).

% Appendix
\appendix
\section{Supplemental cases\label{ApxA}}

\begin{figure}[htb]
    \centering
    \includegraphics[width=1.0\hsize]{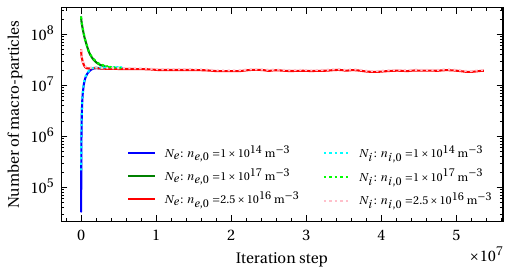}
    \caption{\miold{Case modeling IMPF reactor.
    A PIC simulation snapshot of the electron density profile of the filamentary magnetized plasma within a $23.2~\text{mm}\times11.6~\text{mm}$ 2D rectangle CCRF discharge container, which opens a symmetric pair of 0.2-mm-wide vacuum cutout between 19.4-mm-wide inner RF electrode bottom edge and 1.8-mm-wide outer grounded metal bottom edges.}
    \minew{Comparison of different initial states. The curves depict the temporal variations in the numbers of macro-particles throughout the simulations. These three simulations is initialized with $n_{p,0}$ of \SI{1.0e14}{m^{-3}}, \SI{2.5e16}{m^{-3}} and \SI{1.0e17}{m^{-3}}, with the same particle weight $w=73728$. The dashed curves represent the number of $\text{Ar}^{+}$ macro-particles, and the solid curves represent the number of $\text{e}^{-}$ macro-particles. The time step is \qty{18.7}{\ps}.}}
    \label{Fig6}
\end{figure}

\begin{figure}[htb]
    \centering
    \includegraphics[width=1\hsize]{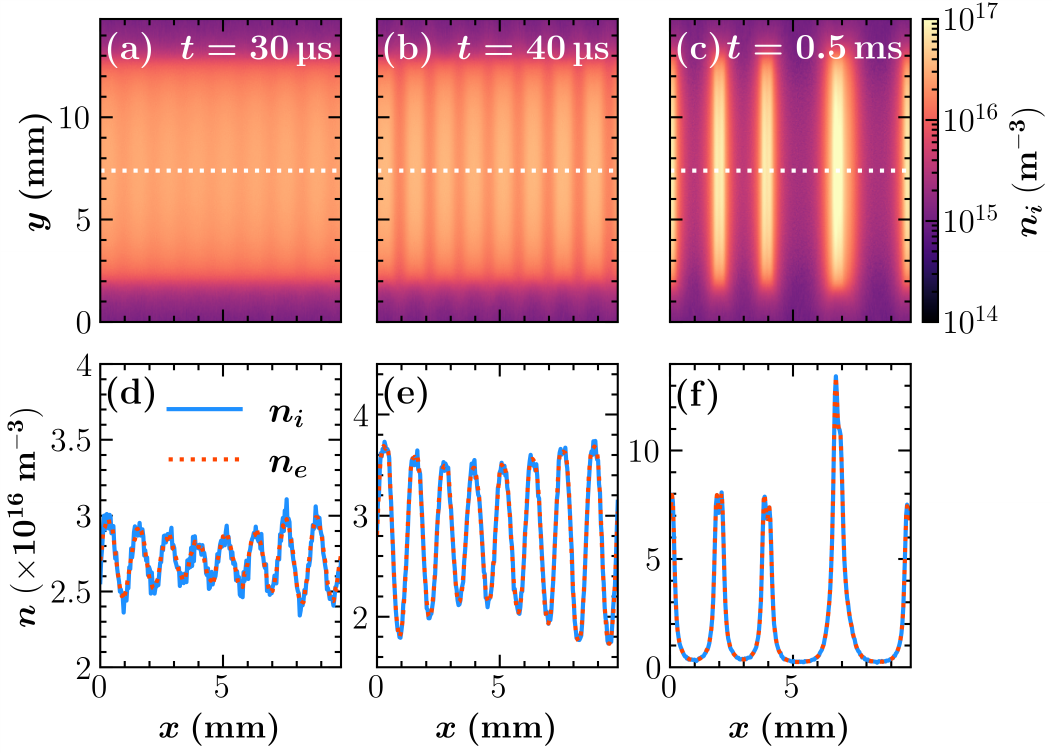}
    \caption{Case with SEE. Plasma population evolution within 500 microseconds. The $\text{Ar}^+$ 2D profiles present plasma distribution at distinct moments: $t = $ (a) \qty{30}{\us}, (b) \qty{40}{\us}, (c) \qty{500}{\us}. The curves plotted in (d), (e) and (f) are the species density varying spatially along the waistline corresponding to (a--c), respectively. Typical modulational feature is presented upon the spatial waveform in (d) as well as (e).}
    \label{Fig7}
\end{figure}

Authors have comprehensively analyzed the performance of argon plasma in different kinds of configurations. Plasma reactor models of varying size, different RF power supplies, effects of surface electron emissions (SEE)\minew{, different initial plasma density} and varying axial magnetic field strength, have been compared. Two valuable supplement\miold{al case}s are mentioned here to support our arguments. They are a \miold{case modeling IMPF type of reactor (see figure~\ref{Fig6})}\minew{comparison of different initial states (figure~\ref{Fig6})}, and a case adding SEE effects (figure~\ref{Fig7}), separately.

\minew{
    We initiated our simulations with different initial densities, \SI{1.0e+14}{m^{-3}}, \SI{2.5e+16}{m^{-3}} and \SI{1.0e+17}{m^{-3}}, and the convergence curves are shown in figure~\ref{Fig6}. These simulations converged to nearly the same micro-particle number, demonstrating the relative independence of the final filamentary regime from the initial plasma density. This is because when the plasma discharge reaches equilibrium, the balanced plasma density and temperature are no longer relevant to the initial settings; and in our simulations, filamentation appears to commence after this equilibrium is stabilized.
}

\miold{
    Figure~\ref{Fig6} presents a snapshot of $n_e$ profile within the rectangle simulation with a IMPF-type reactor immerged in longitudinal external magnetic field ${|\bf B|}=1.3~\text{T}$, which has a size $23.2~\text{mm}\times11.6~\text{mm}$. As mentioned in report~\cite{konopka_2005_complex}, the CCRF reactor chamber of IMPF has a bottom RF electrode with grounded outer loop, separated by a thin gap, and a top grounded metal plate electrode. Sketched in figure~\ref{Fig6}, current 2D configuration setup opens a symmetric pair of 0.2-mm-wide vacuum cutout between 19.4-mm-wide inner RF electrode bottom edge and 1.8-mm-wide outer grounded metal bottom edges. Besides, the physic parameters are the same in section~\ref{Sec2}.
}

\miold{
    Despite a smaller simulation domain referring to the IMPF device, simulation produces the same filament patterns. The experimental observations show that the transversal sheath structures are formed over the vacuum gaps. Remarkably, the bright lumps near both sides are the transversal diffusion residues induced by the periodic boundaries rather than the uninterpretable modeling error. In comparison to the model illustrated in figure~\ref{Fig1}(a), the current reactor configuration is considered to introduce solely the RF alternating electric field above the vacuum gap, serving as a transverse perturbation pumping source. Initially, it was widely believed that this transverse RF electric field was the primary perturbation source leading to the filamentation phenomenon. Consequently, patterns were thought to originate from positions where the transverse RF field is strongest, i.e., the outer edge of the RF electrode~\cite{schwabe_2011_pattern,menati_2019_filamentation}. However, the simulation results indicate that the generation of filamentation does not depend on an externally applied transverse perturbation source, and the thermal perturbations alone are sufficient to elicit the instability.
}

Figure~\ref{Fig7} encapsulates the simulation results incorporating the SEE effects. This case differs only in the engagement of secondary electron emissions. The other simulation parameters are the same in section~\ref{Sec2}. The effect of enabling SEE is to slightly increase the electron density \miold{on the surface of the metal electrode}\minew{near the electrode}, thereby maintain the the quantity of charged particles. In the SWG stage (figures~\ref{Fig7}(d--e)), the mean plasma density at half-height level reaches \qty{2.7e16}{m^{-3}} (35\% higher than mean density in figures~\ref{Fig2}(d--e)), however, the \miold{spatiotemporal distribution}\minew{structural} characteristics of the physical parameters in both sets of the cases remain consistent. \miold{The simulation extends over a time span of \qty{0.5}{ms}, and the evolution trend of the total particle count within the simulated space closely approximates the curve depicted in figures \ref{Fig2} and \ref{Fig4}.} The filamentary pattern observed in the ion density distribution at three selected moments \miold{does not }exhibit \miold{fundamental distinctions in general characteristics of the filamentation compared to the scenario}\minew{limited differences to the cases} without SEE effects. \miold{The density profile on the waistline vividly highlights the modulation of plasma waves. In summary, it can be concluded that}\minew{To conclude,} the presence of SEE effects does not fundamentally impact the filamentation phenomena. \minew{This is consistent with the findings in~\cite{menati_2019_filamentation}.}

\miold{
    \section{Discussions on common electromagnetic effects\label{ApxB}}
}
\miold{
    At high frequencies, electromagnetic effects such as standing waves and skin effects can be observed, which cannot be described using the conventional electrostatic analysis. For a discharge with circular plates (radius R), the standing waves are due to a surface wave that propagates radially into the discharge~\cite{lieberman_2002_standing}. But in current case, the standing wave effects are small, because the radical wavelength $\lambda$ in the discharge plasma is much larger than the radius of the circular plate electrodes of IMPF and MDPX, R = 5 or 8 cm respectively. $\lambda \approx \lambda_0 / \sqrt{1 + \min \{ H, \delta_p \} / 4 h_s} \gg R$, where $\lambda_0 = c / f_{RF} = 5~\text{m}$ is the vacuum wavelength, $\delta_p \sim c / \omega_{pe}$ is the plasma skin depth, $H$ is the bulk plasma thickness of a few centimeters and $h_s$ is sheath thickness of a few millimeters. Skin effects are merely found at high densities where the plasma interior shields itself from the applied fields, and are negligible since the plasma skin depth $\delta_p \gg \sqrt{H R}$. These criteria are provided in the textbook~\cite{lieberman_2005_principles}.
}

% References
\vskip1.5em\section*{References}\bibliographystyle{iopart-num}
\bibliography{Refs}

\end{document}

\title[Supplemental Material]{Supplemental Material: Animated RF dynamics of \\Particle-In-Cell simulations of the filamentation process in magnetized radio-frequency plasmas}

\author{Huidong Huang}%~\orcidlink{0009-0005-8959-8751}}
    %\email{huanghd26@mail2.sysu.edu.cn}
    \affiliation{
     Sino-French Institute of Nuclear Engineering and Technology, Sun Yat-Sen University, Zhuhai 519082, People's Republic of China}
\author{Jian Chen}%~\orcidlink{0000-0001-9807-489X}}
    \email{chenjian5@mail.sysu.edu.cn}
    \affiliation{
     Sino-French Institute of Nuclear Engineering and Technology, Sun Yat-Sen University, Zhuhai 519082, People's Republic of China}
\author{Zhibin Wang}%~\orcidlink{0000-0002-6812-7855}}
    \email{wangzhb8@sysu.edu.cn}
    \affiliation{
     Sino-French Institute of Nuclear Engineering and Technology, Sun Yat-Sen University, Zhuhai 519082, People's Republic of China}
\date{\today}

% Abstract & Keywords

\maketitle

% Main Body
Movie S1. Animation of the RF dynamics of the Standing Wave Growing stage.

Movie S2. Animation of the RF dynamics of the Pattern Growing stage.

% Acknowledgements

% Appendix

% References
\bibliography{Refs}